\documentclass[useAMS,usenatbib]{mn2e}
\usepackage{amssymb,amsmath,psfig,times}
\usepackage{graphicx}

\newcommand{\sw}{\textit{Swift}}

\newcommand{\lx}{$L_{\mathrm{BAT}}$}
\newcommand{\lr}{$L_{\mathrm{AT20}}$}

\newcommand {\gsim}{ \lower .75ex \hbox{$\sim$} \llap{\raise .27ex \hbox{$>$}} } 
\newcommand {\lsim}{ \lower .75ex\hbox{$\sim$} \llap{\raise .27ex \hbox{$<$}} }

\newcommand\apj{ApJ}
\newcommand\aap{A\&A}

\newcommand\mnras{MNRAS}
\newcommand\nat{Nature}
\newcommand\apjl{ApJL}
\newcommand\apjs{ApJS}

\newcommand\aj{AJ}

\title[The AT20G view of \sw\,/BAT selected AGN]{The AT20G view of \sw\,/BAT selected AGN: high frequency radio waves meet hard X--rays}
\author[D. Burlon et al.]
{D. Burlon\thanks{E-mail: dburlon@physics.usyd.edu.au}$^{1,2}$,
G. Ghirlanda$^{3}$, 
T. Murphy$^{1,2}$,
R. Chhetri$^{4,5}$,
E. Sadler$^{1,2}$, 
and M. Ajello$^{6}$\\
$^{1}$Sydney Institute for Astronomy, School of Physics, The University of Sydney, NSW 2006, Australia\\
$^{2}$ARC Centre of Excellence for All-sky Astrophysics (CAASTRO), The University of Sydney, NSW 2006, Australia\\
$^{3}$INAF, Osservatorio Astronomico di Brera, via E. Bianchi 46, I-23807 Merate, Italy\\
$^{4}$ Australia Telescope National Facility, CSIRO Astronomy and Space Science, PO Box 76, Epping, NSW 1710, Australia\\
$^{5}$ Department of Astrophysics and Optics, School of Physics, University of New South Wales, NSW 2052, Australia\\
$^{6}$ Space Sciences Laboratory, 7 Gauss Way, University of California, Berkeley, CA 94720-7450, USA}

\begin{document}
\date{Accepted 2013 February 21. Received 2013 February 20; in original form 2012 December 16}
\pagerange{\pageref{firstpage}--\pageref{lastpage}} \pubyear{2002}
\maketitle \label{firstpage}

\begin{abstract}
We cross--matched the six--year \sw/Burst Alert Telescope (BAT) survey of Active Galactic Nuclei (AGN) with the AT20G radio survey of the southern sky, which is one of the largest high--frequency radio surveys available. With these data we investigated the possible correlation between the radio and the X--ray emission at the highest radio and X--ray frequencies. We found 37 AGN with a high probability of association ($>80$\%), among which 19 are local Seyfert galaxies (with median redshift $z=0.03$) and 18 blazars.  
We found that $\approx 20$\% of the AGN detected in hard X--rays are also bright radio sources at 20~GHz, but the apparent correlation between the radio and hard X--ray luminosity is completely driven by the different median redshifts of the two subgroups of AGN.
When we consider only the local Seyfert sample we find no evidence of a correlation between their 20~GHz and 15--55 keV power. Therefore it appears that at high frequencies the radio--X connection, which had been previously observed at lower frequencies, disappears. The disappearance of the radio--X correlation at high radio and X--ray frequencies could be tested through Very Long Baseline Interferometry and the use of the Nuclear Spectroscopic Telescope Array (NuSTAR) satellite.

\end{abstract}

\begin{keywords}
radiation mechanisms: non-thermal, quasars: general, radio continuum: general, X--rays: general\end{keywords}

\section{Introduction}
Active Galactic Nuclei (AGN) are among the most powerful sources of radiation across the whole electromagnetic spectrum, from $\gamma$-rays (up to TeV energies in some cases) down to radio waves. Accretion onto a central supermassive black hole (SMBH) is generally accepted to be one of the dominant physical mechanisms that power these systems. Another source of energy is believed to be the rotational energy of the SMBH itself, which in some cases can exceed the gravitational energy extracted through accretion.

The radiation output of quasars and Seyfert galaxies peaks in the ultraviolet \citep{shakura73}, producing the so-called ``big blue bump'' in the spectral energy distribution (SED), while only a fraction of the energy available through accretion is reprocessed into high energy emission (from keV energies to hundreds of keV)  via thermal comptonization, which produces a second, less dominant peak in the SED, and which is generally thought to cause the cosmic X--ray background \citep[CXB,][]{giacconi62}. The most accepted scenario for  comptonization requires a population of hot electrons \citep[i.e. a corona, see][]{haardt91} that ``sandwiches'' the inner part (tens of gravitational radii $R_g = GM/c^2$) of the accretion disk, and which is kept hot by magnetic reconnection. Additionally, an outflow or a jet could contribute to the total emitted power, more prominently at radio frequencies through synchrotron radiation and less markedly at $\gamma$--ray frequencies \cite[see e.g.][]{teng11, lenain10}.

When AGN also produce relativistic jets which are pointing in the direction of the observer, they are referred to as blazars. In this subclass of AGN the SED shows two clear peaks, or ``humps''. In leptonic models, all the radiation we see coming from the central object is generally thought to be generated by a single population of electrons, i.e. directly or indirectly by the jet itself. This is achieved via synchrotron emission producing the lower--energy hump and via a combination of external compton (EC) and Synchrotron--self--Compton (SSC) producing the high energy hump. Only in a handful of cases the thermal radiation coming from the accretion disk is powerful enough to ``pierce'' through the synchrotron hump, and  could be used to estimate the mass and accretion rate of the AGN. All blazars are powerful radio sources, and the relative contribution of the two peaks in the SED as well as the position of their synchrotron peaks classify them into subclasses of radio (and total) power \citep{fossati98}.

In practice, a popular method of classifying AGN uses the ratio of the radio (at 5~GHz) to optical (at 4400~\AA) fluxes, through the so--called radio-loudness parameter $R=F_{radio}/F_{optical}$. Initially, the distinction between radio-loud ($R>10$) and radio-quiet ($R\lsim10$) objects  appeared sharp \citep{strittmatter80, kellermann89, kellermann94} if derived from the distribution of the $R$ parameter of AGN present in the Palomar Bright Quasar Survey. Up to five other similar proxies of R have been defined in the literature \citep[for a recent review see][and references therein]{elvis12}.
The existence of a bimodality has been ever since debated in the literature in favor \citep[see e.g.][]{miller90,stocke92,hooper95,ivezic02,jiang07} or against it \citep[see e.g.][]{condon80, white00, cirasuolo03, rafter09}. A possible source of confusion is the different size, sensitivity limits, cuts in redshift, optical and radio frequencies used to classify AGN through the R parameter.

Recently \cite{mahony12} used a soft X--ray selected sample of several hundreds quasars and did not find any evidence of bimodality. Seyfert galaxies are traditionally considered radio-quiet sources but, for example,  \cite{ho_peng01} showed that considering their nuclear luminosities, most Seyfert galaxies are radio-loud.

To account for this dichotomy, it was proposed that just rapidly spinning black holes can produce relativistic jets responsible for non--thermal emission \cite[e.g.][]{blandford90}. Nonetheless, in later years the evidence for high spin came from a radio--quiet object (MCG-6-30-15, \citealp{tanaka95}). 
Other works \citep{nandra07, patrick12} showed that broad relativistic lines are frequently observed in Seyfert galaxies and in a handful of cases (e.g. NGC~3516, NGC~7469) they could also constrain the spin. 
Other results are at odds with the idea of radio--quiet objects hosting slowly spinning black holes. One example is the high efficiency required by accretion in order to account for the integrated CXB \citep{elvis02}, which is more easily explained if the BH are spinning. Other examples are X--ray observations that indicate that the emission extends well within the innermost stable orbit of non--rotating black holes \citep[e.g.][]{fabian02, iwasawa96}. 
Moreover, both Seyfert galaxies and low--luminosity AGN (LLAGN) were shown to be radio emitters at some level. For instance, \cite{ho_ulvestad01} have shown that 85\% of Seyfert AGN are detected at 5~GHz, with a wide range of radio powers and morphologies, where structures could be resolved by very long baseline interferometry (VLBI). 
\cite{nagar02}, \cite{ulvestad03} and \cite{giroletti09} showed clear structures (blobs or knots) at the sub--parsec level. 
The origin of radio emission at this level is not yet fully understood. In most cases it is 
associated to synchrotron emission just as in the radio--loud case \citep{giroletti05,ishibashi10} and, if so, should highlight the presence of a jet or outflow also in this class of objects. Alternatively, thermal free--free emission is viable option \citep[see e.g.][for the radio emission of NGC~1068, among others]{gallimore04}.

The radio--X connection has been  addressed by different authors, but given the different energy ranges used in the high--energy band coupled with different radio frequencies, the results are often discordant (and will be discussed in Sect.~4). For instance, in the seminal work by \cite{merloni03} a relation between the  radio luminosity at 5~GHz and the X--ray 2--10 keV luminosity $L_R \propto L_X^{0.6}$ (i.e. $L_X \propto L_R^{1.7}$) was found and  a solid theoretical interpretation was discussed in relation to the scale invariance of black hole jets \citep[see also][]{heinz03}. Other authors find a slope close to unity \citep{canosa99, brinkmann00, salvato04, panessa07} or a completely different slope $L_X \propto L_R^{0.6}$ \citep{wang06}  similar to what observed in few Galactic black holes \citep{corbel13}. 
While completing this paper, \citep{bonchi13} reported an updated version of the Fundamental Plane, with a slope even harder than the original one: $L_R \propto L_X^{0.39}$.   

The aim of this work is to study \emph{for the first time} the radio--X connection in local Seyfert galaxies in the hard--X ray, high--frequency radio regime.  We want to test whether the correlation between the radio and X--ray power at high radio and X--ray frequencies is as strong as at lower frequencies, or conversely tends to vanish (as it seems to be the case when high resolution  Very Long Baseline Interferometry observations are used - Panessa \& Giroletti, 2013, MNRAS submitted). To these aims we cross--match the sources present in the BAT survey (15--55 keV) with the sources detected by the high frequency (20~GHz) Australia Telescope Compact Array (ATCA) survey.
A similar  complementary work is in progress \citep{panessa11} adopting the sample of AGN detected by the INTEGRAL satellite, and the 1.4~GHz radio luminosity.
Although our primary focus is on Seyferts, we also consider the blazars found through the cross--matching, in other to investigate the possible gap in the radio--to--gamma ray connection in blazars recently studied  \citep{mahony10, ghirlanda10b,ackermann11}. 

Another approach to the present work is to investigate whether the correlations arise when long--term averages are used, or conversely if there is a radio--X connection at all scales. When at lower radio frequencies the ``core'' luminosities (albeit the ``core'' could mean jet structures of the order of kpc) are confronted to the high--energies the correlation holds both for pointed and survey (i.e. averaged) X--ray observations. Choosing higher radio--frequencies allow us to be less contaminated by the extended structures and therefore a shorter (but still somewhat averaged) radio output. We expect that if we compared hard X--rays from a survey telescope with lower radio frequencies \citep[see an example in][]{panessa11}, we would still recover a strong correlation, possibly because we would be confronting longer term averages, instrumental/observational in the case of X--rays and physical in the case of radio waves. 
 Nevertheless, we note that there is a still large difference in the time scale for X--ray emission (years, for all--sky surveys) and radio waves (several thousand years in kpc--scale regions).
The best test we envisage for the future will be to investigate the radio--X connection using focussing hard X--ray telescopes like NuSTAR \citep{harrison10} and VLBI radio luminosities, which should both probe at the same time the very inner part of the AGN.

In Sect.~2 we briefly discuss the two surveys and the cross--correlation technique. We present our results in Sect.~3 and discuss them  in Sect.~4. We summarise our findings in Sect.~5. A standard flat Universe with $h=\Omega_{\Lambda}=$0.7 is assumed.

\section{The sample}
We give a brief description of the two blind surveys we used for this work, but we refer the reader to the respective original publications for a detailed discussion of both surveys' strategies and peculiarities. In \S~\ref{sect:associations} we describe the cross--correlation technique we adopted.

\subsection{The \sw/BAT survey}
The Burst Alert Telescope \citep[BAT;][]{barthelmy05} instrument on--board the \sw\ satellite \citep{gehrels04} is a coded mask instrument originally thought to alert the X--ray and Optical--UV cameras to follow-up $\gamma$--ray burst afterglows. While surveying the sky waiting for a trigger, the BAT continuously monitors up to 80\% of the sky thanks to its wide field of view ($120^{\circ} \times 90^{\circ}$ partially coded). After several years of deep exposure, the BAT survey covers almost uniformly the whole sky down to the limiting sensitivity. In our analysis we used the data accrued in 6 years  of exposure, down to a sensitivity limit $\sim 6\times10^{-12}$ erg/cm$^2$/s. The BAT survey is presently \citep[see][for the latest release]{ajello12} the deepest blind scan in the sky above 10~keV making it the optimal candidate for the goal of confronting the hard X--ray and radio properties of AGN.
\begin{figure}
\hspace{-0.5cm}
\includegraphics[width=9cm]{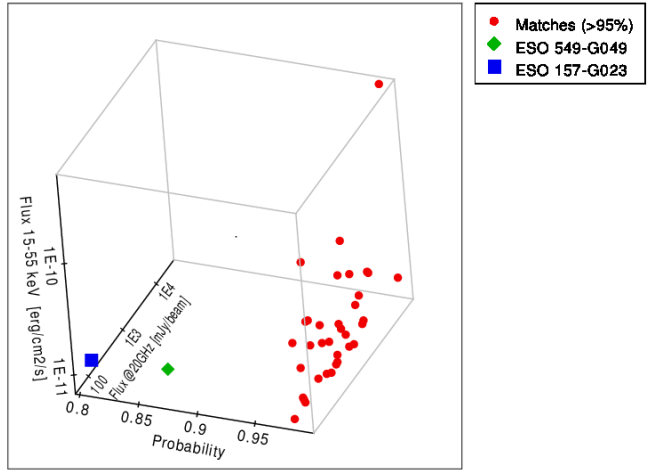}
{\caption{Association (posterior) probability between the \sw/BAT and AT20G samples versus radio flux density and hard X--ray flux of the matched sources. Two `outliers' are clearly visible in the low-flux low-probability section of the data cube. These are the only two AGN for which the association probability is between 80\% and 95\%. }
\label{probass} \vfil} 
\end{figure}
Data screening and processing were described in details in \cite[][A08a]{ajello08a} and \cite[][A08b]{ajello08b} but we summarise hereafter the main characteristics of the survey. The chosen energy interval for (conservatively) computing the fluxes was 15--55 keV, while the spectra are traditionally extracted in the 15--200 keV energy range. The all--sky image was obtained by averaging the (weighted) per--pointings, resulting in a final sample of $\approx 350$ AGN, comprising mostly Seyfert--like galaxies (284) and blazars. 
Just $\sim5\%$ of the sources of the BAT catalogue still escape  identification, and this incompleteness is independent of the Galactic latitude \citep[see e.g.][]{ajello12}.
Each of these sources has a signal--to--noise ratio $SNR \gsim 5\sigma$, and was identified by cross--matching it against the previous catalogues of \cite{tueller08}, \cite{cusumano10}, \cite{voss10}, and \cite{burlon11}. Whenever possible, optical identifications were taken from \cite{masetti08, masetti09, masetti10} as described in \cite{ajello12}. For the blazars we checked their classification against the BZCAT\footnote{http://www.asdc.asi.it/bzcat/} catalogue \citep{massaro09}.

\subsection{The AT20G survey}
The AT20G survey \citep{murphy10} is one of the largest blind surveys at high radio frequencies ever conducted from ground. It covers $\sim 20,000$ deg$^2$ to a limiting flux density of 40 mJy/beam. The key feature of the survey was its two--phase strategy, where a fast blind scan (possible due to the fast--scanning mode of the Australian Telescope Compact Array) was followed by regular snapshots by the ATCA.

\begin{figure}
\includegraphics[width=\columnwidth]{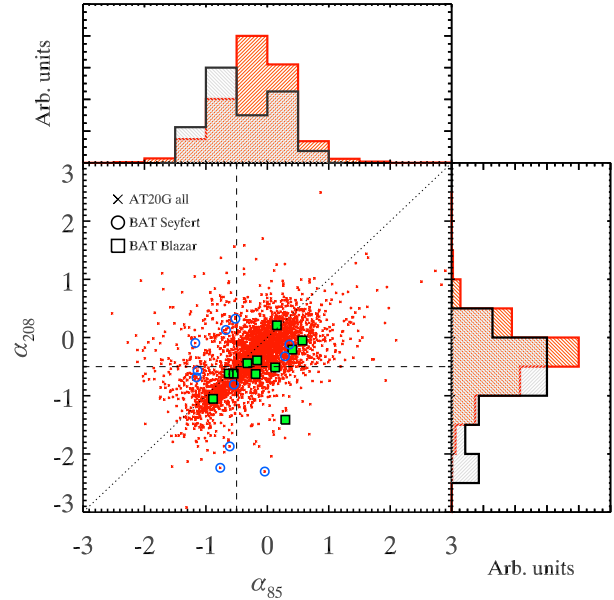}
{\caption{Scatter plot of the radio spectral indices of the all AT20G sources are shown in red crosses. Seyfert galaxies (blazars) are shown with circles (squares). The two panels show the (arbitrarily rescaled) projected distributions along the respective axis of the whole AT20G population (red histogram) and the whole cross-correlated subsample (black histogram).}
\label{alfalfa} \vfil} 
\end{figure}

In the first phase the ATCA scanned the whole \emph{southern} sky at 20~GHz, with a speed of $15^{\circ}$~min$^{-1}$ in declination at the meridian. The scanning strategy exploited the Earth rotation, while sweeping regions of the sky $10^{\circ}-15^{\circ}$ wide in declination. The rms sensitivity of the first phase was $\simeq 10$~mJy. In the second phase, the sources brighter than $5\sigma$ where followed--up with an hybrid array configuration and two 128 MHz bands, combined during data processing at the central frequency of 19.904~GHz (that we refer to as 20~GHz throughout the paper). The sources with declination~$< -15^{\circ}$ were observed also at 4.8~GHz and 8.64~GHz (which we refer to as 5~GHz and 8~GHz), with an east--west configuration. The exclusion of sources with declination $> -15^{\circ}$ for the lower frequencies is motivated by the poor $(u,v)$ coverage of east--west arrays near the equator. 

\begin{figure*}
\includegraphics[width=\columnwidth]{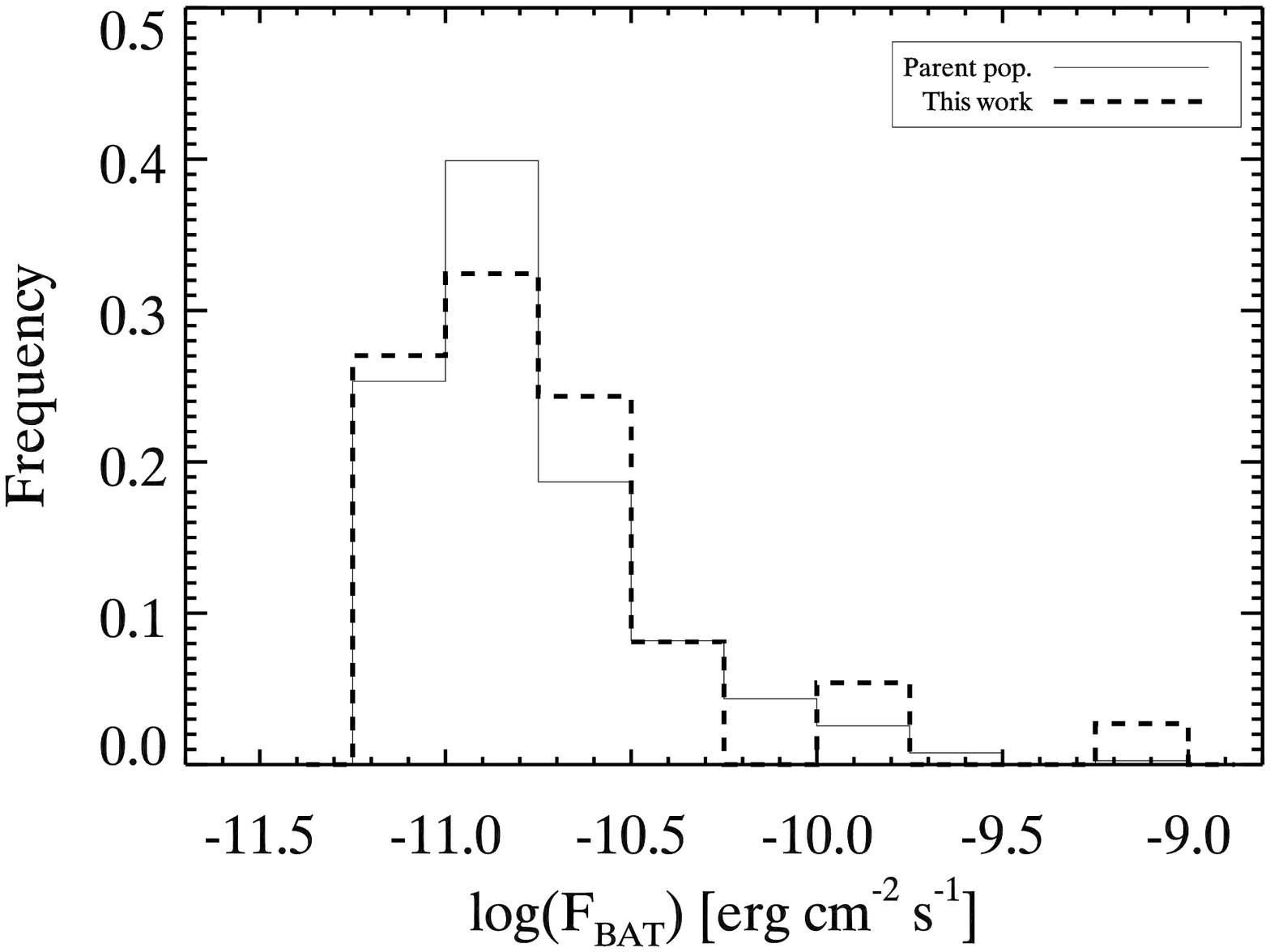}
\includegraphics[width=\columnwidth]{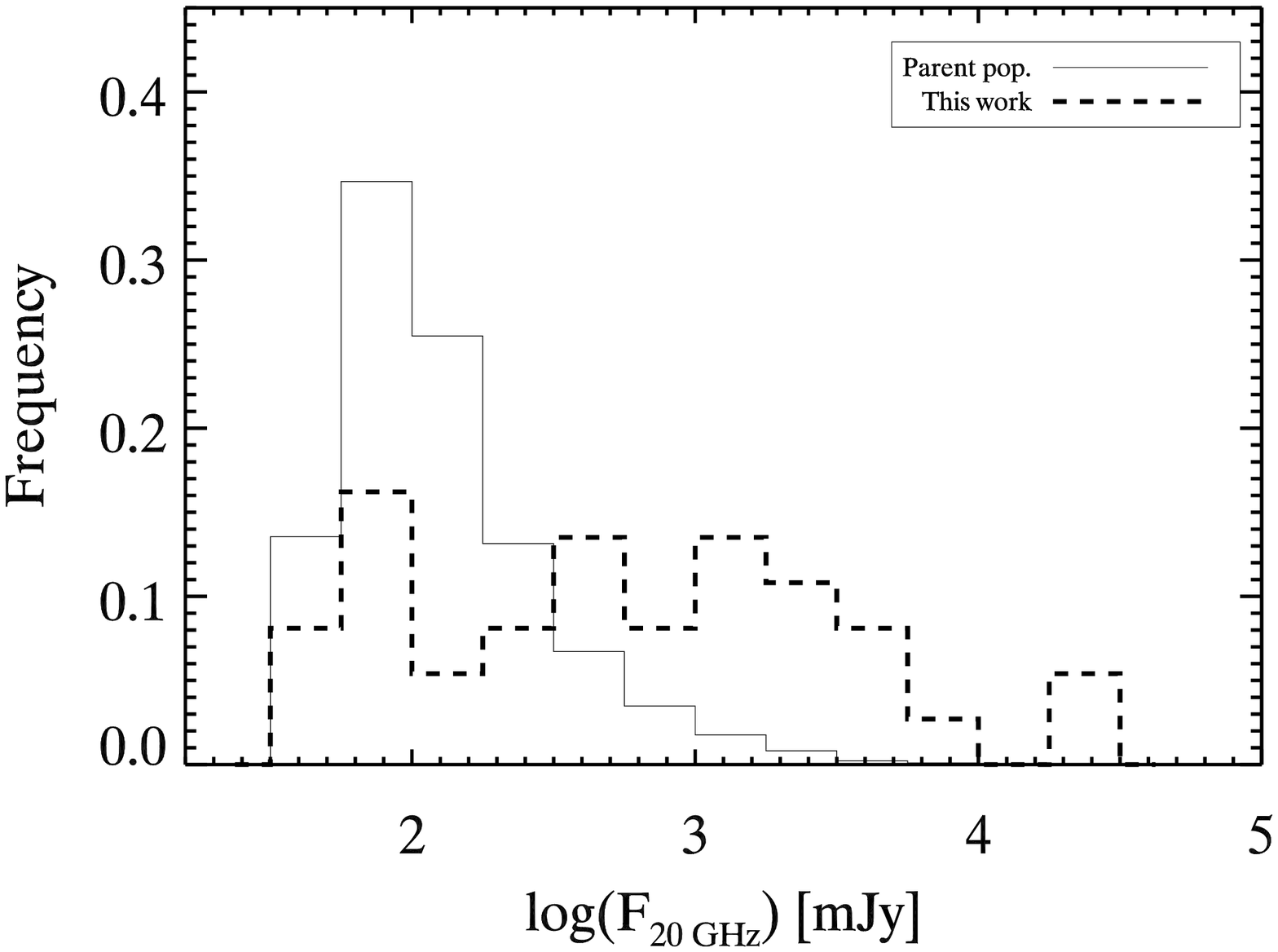}
{\caption{Distribution of the parent populations flux, compared with the associated AGN. The distributions have been rescaled. 
\textit{(Left:)} \sw/BAT flux in the energy band 15--55 keV for the whole 6--year catalog (thin line) and the associated subsample (thick dashed line). 
\textit{(Right:)} AT20G flux densities at 20~GHz, with the same notation as left panel.}
\label{ksflux} \vfil} 
\end{figure*}

\subsection{Source Associations} \label{sect:associations}
For the positional cross--matching of the BAT and AT20G source catalogs we adopt a Bayesian method already used in cross--matching the \textit{Fermi} AGN with the AT20G survey \citep{ghirlanda10b} and first developed for the identification of the {\it Fermi} sources \citep[][and references therein]{abdo10a}. 
This method, given the BAT source position and accuracy (which we assume to be half of the point spread function) assigns a probability to every radio source found within the BAT error circle. This probability is computed accounting for the surface density of radio candidates around each BAT source. To this aim, the prior probability of the radio sources (i.e. the confidence that the radio catalogue contains the counterpart of the X--ray selected AGN) is calculated as per appendices B and C of \cite{abdo10a}, and is assumed equal for all the sources. In practice, this translates into calculating the average number of false associations between the AT20G catalogue and 100 randomly generated mock X--ray catalogues.
{Similarly to what done in  \cite{abdo10a} and \cite{ghirlanda10b}, we call an `association' a unique match (i.e. a single radio source within the BAT error circle) with a posterior probability greater than 80\%. 
We comment below on some of the sources that are flagged in the original paper by \cite{murphy10}. While this method allows us to derive a figure of merit for the associations found in the two catalogs, we note that by only searching for positional matches (i.e. radio sources within the BAT error circle), we could derive a similar --albeit smaller-- number of associations.



\begin{table*}
\caption{Cross--matched sample. The sources are listed with increasing AT20G name. The \sw/BAT flux is in units of erg/cm$^2$/s, while the radio flux densities are in mJy. The BAT fluxes have been de--absorbed with the recipe of Burlon et al. (2011) where log$N_H$ was in excess of 23.5, and these sources are marked in boldface accordingly. References are in the text. } 
\label{tabellon}
\begin{tabular}{@{}lllllllllllll@{}}
\hline
\hline
Name 							& Type 		& AT20G Name 		& Redshift 	& Prob. & Flux$_{15-55}$ 		& S20 & 			$\sigma_{S20}$ & S8 	& $\sigma_{S8}$ 		& S5 & $\sigma_{S5}$ &  $\Gamma_{BAT}$\\
								&			&					&			&		& [cgs]					& [mJy]		& [mJy]		& [mJy]		& [mJy]				& [mJy]		& [mJy]	&					\\
\hline
{\bf NGC 612}                        & Sy2        & J013357-362935 &      0.0297&       0.994 &    3.535e-11 &       440  &       \dots &     \dots &        \dots &           \dots &           \dots &        1.63 \\
NGC 1052                       & Sy2        & J024104-081520 &     0.0050&       0.952 &    1.359e-11 &       1199 &       	59 &       \dots &        \dots &           \dots &           \dots &    1.47 \\
{\bf NGC 1068}                       & Sy2        & J024240-000046 &     0.00378&       0.995 &    9.985e-10 &       474  &      	17 &       \dots &        \dots &           \dots &           \dots &    2.23 \\
PKS 0312-77                    & FSRQ       & J031155-765151 &       0.2230&       0.987 &    9.237e-12 &       1238 &     81 &       1292 &    	  67  &      		 924  &       		46  &   2.08 \\
RBS 0471                       & Sy1        & J034730-303521 &      0.095&       0.979 &    7.383e-12 &       55  &       	2 &       42 &     	  2 &       				57  &       		3 &       		 2.04 \\
IGR J03532-6829                & BLAZ       & J035257-683117 &      0.0869&       0.987 &    9.778e-12 &       68  &       3 &       114 &      	 4 &       			165  &      			 5 &        2.30 \\
ESO 549-G049                   & Sy2        & J040314-180953 &      0.0262&       0.874 &    1.394e-11 &       54  &       4 &       106 &      	 7 &       			146  &       			7 &        1.75 \\
RX J0405.5-1308                & FSRQ       & J040534-130813 &       0.571&       0.960 &    8.180e-12 &       1576 &      77 &       \dots &        \dots &           \dots &     	     \dots &        2.14 \\
ESO 157-G023                   & Sy2        & J042203-562127 &      0.0433&       0.800 &    9.956e-12 &       95  &       5 &       105 &      	 5 &       			85  &     		 	4 &        1.67 \\
LEDA 177818                    & Sy2        & J044436-280914 &       0.147&       0.983 &    9.755e-12 &       221  &      10 &       1431 &     	 19  &      			 2245 &     		  46  &        2.17 \\
Pictor A                       & Sy1        & J051949-454643 &      0.0351&       0.985 &    2.136e-11 &       6320 &      110 &       \dots &        \dots &           \dots &           \dots &        1.90 \\
ESO 362-G021                   & BLLAC      & J052257-362730 &      0.0553&       0.983 &    1.612e-11 &       3909 &      156 &       6566 &     	244  &      			 9066 &     		  342  &        1.85 \\
PKS 0524-460                   & FSRQ       & J052531-455754 &        1.479&       0.988 &    9.642e-12 &       483  &     24 &       698 &     	  35  &     			  845  &    		   42  &        1.22 \\
SWIFT J0539.9-2839             & FSRQ       & J053954-283956 &        3.103&       0.986 &    1.009e-11 &       677  &     44 &       1143 &    	   60  &      		 1278 &      		 64  &        1.55 \\
PKS 0558-504                   & Sy1NL      & J055947-502652 &       0.1369&       0.983 &    1.014e-11 &       78  &      4 &       70 &       	4 &       			104  &      	 		5 &        2.04 \\
PKS 0637-752                   & FSRQ       & J063546-751616 &       0.6350&       0.988 &    1.079e-11 &       3142 &     206 &       4358 &     	  227  &      		 4804 &     		  240  &        1.92 \\
3C 206                         & Sy1        & J083950-121434 &       0.1978&       0.995 &    1.216e-11 &       323  &     16 &       \dots &        \dots &           \dots &           \dots &        2.04 \\
PKS 0921-213                   & BLAZ       & J092338-213544 &      0.05299&       0.989 &    9.404e-12 &       328  &     22 &       390 &      	 20  &       		307  &       		20  &        2.24 \\
QSO B1127-145                  & FSRQ       & J113007-144927 &        1.187&       0.990 &    1.880e-11 &       1865 &     91 &       \dots &        \dots &           \dots &           \dots &        1.75 \\
3C 279                         & FSRQ       & J125611-054721 &       0.5361&       0.995 &    1.233e-11 &       20024 &    996 &       \dots &        \dots &           \dots &           \dots &        1.50 \\
{\bf NGC 4945}                       & Sy2        & J130527-492804 &     0.00187&       0.995 &    5.263e-10 &       726  &      36 &       786 &      	 39  &       		1566 &      		 78  &        1.47 \\
IGR J13109-5552                & Sy1        & J131043-555209 &       0.1040&       0.991 &    1.219e-11 &       47  &      1 &       321 &      	16  &       			329  &      			 16  &        1.40 \\
Cen A                          & Sy2        & J132527-430104 &     0.00182&       0.987 &    5.687e-10 &       28350 &    \dots & 	 	\dots &        \dots &           \dots &           \dots &        1.85 \\
{\bf Cir Galaxy}                     & Sy2        & J141309-652020 &     0.00144&       0.988 &    1.035e-09 &       97  &       5 &       156 &      	7 &       			304  &      			 15  &        2.09 \\
PKS 1510-089                   & FSRQ       & J151250-090558 &       0.3599&       0.970 &    2.656e-11 &       2933 &      142 &       \dots &        \dots &           \dots &     	      \dots &        1.36 \\
PKS 1549-79                    & FSRQ       & J155658-791404 &       0.1495&       0.976 &    1.170e-11 &       821  &      41 &       1978 &    	  99  &      		 3326 &       		166  &        2.47 \\
SWIFT J1656.3-3302             & FSRQ       & J165616-330207 &        2.400&       0.984 &    2.610e-11 &       287  &      14 &       241 &     	12  &       			220  &     		  11  &        1.60 \\
PKS 1830-21                    & FSRQ       & J183339-210341 &        2.500&       0.987 &    2.343e-11 &       5495 &      360 &       \dots &        \dots &           \dots &           \dots &        1.50 \\
NGC 6860                       & Sy1        & J200848-610938 &      0.01488&       0.981 &    3.309e-11 &       64  &      3 &       \dots &        \dots &           \dots &           \dots &        1.98 \\
IC 5063                        & Sy2        & J205202-570407 &      0.0113&       0.986 &    4.034e-11 &       130  &      6 &       230 &      	12  &       			453  &     		  23  &        1.89 \\
QSO B2052-47                   & FSRQ       & J205616-471447 &        1.488&       0.992 &    1.015e-11 &       1171 &      57 &       3806 &    	  199  &    		   3204 &       		160  &        2.25 \\
PKS 2126-158                   & FSRQ       & J212912-153841 &        3.280&       0.987 &    1.268e-11 &       1073 &      54 &       1647 &    	 82  &      			 1528 &      	 76  &        1.62 \\
PKS 2149-306                   & FSRQ       & J215155-302753 &        2.345&       0.987 &    3.526e-11 &       1846 &      89 &       \dots &        \dots &           \dots &           \dots &        1.48 \\
NGC 7213                       & Sy1.5      & J220916-471000 &      0.0277&       0.987 &    2.462e-11 &       123  &      6 &       161 &      	8 &       			136  &      			 7 &        1.92 \\
3C 445                         & Sy1        & J222352-021043 &      0.05639&       0.987 &    2.112e-11 &       60  &      3 &       \dots &        \dots &           \dots &           \dots &        1.99 \\
PKS 2227-08                    & FSRQ       & J222940-083254 &        1.561&       0.987 &    9.945e-12 &       3230 &      154 &       \dots &        \dots &           \dots &           \dots &        1.52 \\
PKS 2356-61                    & Sy2        & J235844-605246 &      0.0963&       0.986 &    9.463e-12 &       311  &      12 &       1482 &    	 21  &     			  2124 &       		41  &        1.80 \\
\hline
\hline
%
\end{tabular}
\end{table*}

We find 37 BAT AGN with a radio counterpart in the AT20G survey. Not surprisingly the blazar class has a detection rate of $\approx 65\%$, which is much higher than the equivalent for Seyfert galaxies, i.e. $\approx 14\%$. These are listed in Table~\ref{tabellon}, along with the posterior probability of the association. About half of the associations (19/37) are classified as Seyfert galaxies, roughly equally distributed into the two subclasses of optically Seyfert ``type 1'' (7, plus 1 Sy1 Narrow Line) or ``type 2'' (11). There are 18 AGN classified as either BL Lacs (1) or FSRQ (15) blazars, and 2 classified as generic blazar type. The absolute majority of the FSRQ class is no surprise, given that our parent sample is the \sw/BAT one, which is well known to pick preferentially this class of AGN. 
Note that all but two Seyfert 2 AGN (i.e. ESO 549-G049 and ESO 157-G023) have a probability in excess of 95\%, and that both the less certain associations are close to the detection limit of both surveys. This is represented in Fig.~\ref{probass}, where the two sources are clearly displaced from the bulk of the associations.
 
\emph{Properties of the cross--matched sample:}
One of the results of the AT20G survey is that a large fraction of sources are not well fit by a single power law even in the small bandpass of the ATCA telescope. In Fig.~\ref{alfalfa} we show the distribution of the spectral index of the several thousands sources of the AT20G for which the spectral index could be  computed \citep[see for comparison][]{murphy10, massardi11}. In the same plot we show the radio spectral indices of the BAT sources associated to the radio counterparts. 
The spectral indices where calculated in the standard way taking from \cite{murphy10} the flux densities ($S$) at frequencies $\nu_1$ and $\nu_2$:
$$ \alpha_{12} = \frac{log_{10}(S(\nu_1)/S(\nu_2))}{log_{10}(\nu_1/\nu_2) } $$
Fig.~\ref{alfalfa} shows that the associations we found do no show a particular spectrum, i.e. they populate all the four quadrants of the plane identified by the dashed lines corresponding to $\alpha = -0.5$ which is often adopted as a separation between hard and soft radio sources. \cite{chhetri12} pointed out that AT20G is a high frequency survey and hence it has a larger fraction of flat--spectrum compact radio sources ($\sim 80$\%) than the surveys at lower frequencies ($\sim 25$\%).

We checked also whether the association method tended to pick the brighter end of the distribution of both populations. In Fig.~\ref{ksflux} we show that albeit this is not true for the hard X--ray population, it seems that the radio part of our association picks up preferentially  brighter sources. 
When we compare the \sw/BAT flux distributions (left panel), it is apparent how this work is sampling AGN which are on average distributed exactly as the parent population. When we look at the 20~GHz flux distributions, the associated AGN have a flat distribution. Note that we chose to compare the flux densities at 20~GHz because this is the most sampled frequency of the radio survey (the follow--up completeness is $>90\%$ only above 100 mJy, see \citealp{massardi11}).
\begin{figure}
\includegraphics[width=\columnwidth]{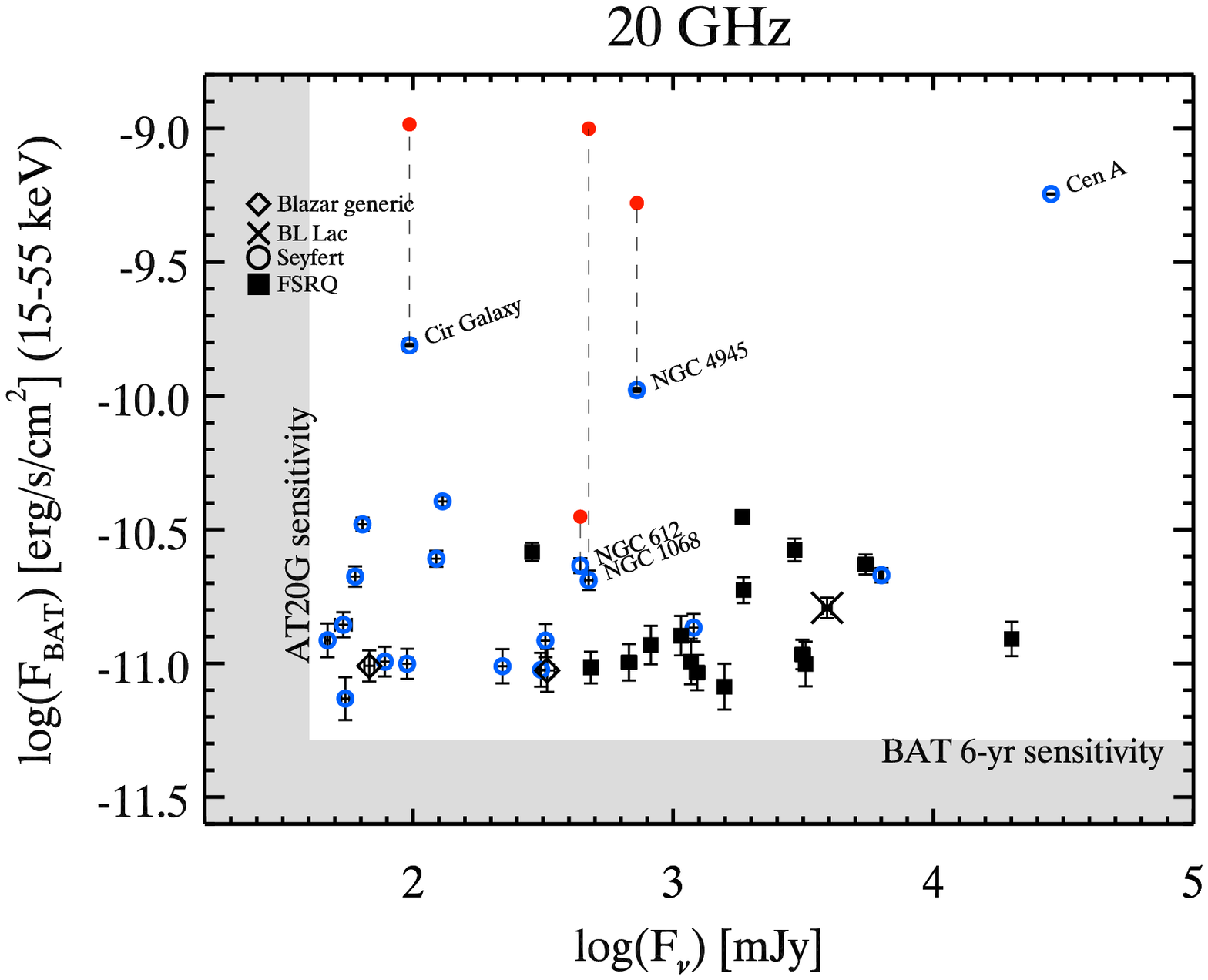}
{\caption{Scatter plot of the \sw/BAT flux in the energy band 15--55 keV vs. 20~GHz flux density. The sensibilities of the two surveys are represented with the grey shaded area. Seyfert galaxies are plotted with blue open (red filled) circles if their flux from the six-year BAT survey (unabsorbed as per \citealp{burlon11}) is used. Blazars are represented with black squares (FSRQ), crosses (BL Lac), and diamonds (unclassified).}\label{20ghz_flux} \vfil} 
\end{figure}

\emph{Notes on `extended' sources\footnote{In the original AT20G catalog 5 of our sources were originally flagged as `poor'. Their flux density measurements are anyway unaffected due to the method (referred to as \textit{Triple Product Method}; see \citealp{murphy10} for additional details) used, which is robust against the effects of decorrelation (phase fluctuations of the visibilities). These flagged data have no significant effect for the analysis.}:} 
\cite{massardi08} reported that the radio galaxy Fornax~A (that is not present in our sample) is the only bright extragalactic source known to be missing from the AT20G. This incompleteness is a result of its 20~GHz flux coming from the lobes being resolved by the ATCA beam. In our sample 6 sources (2MASX~J04450628--2820284, NGC~1068, Circinus, PKS 2356--61, IGR J03532--6829, ESO--362-G021) are flagged as 'extended', while 3 additional ones (Cen~A, Pictor~A, NGC~612) are additionally flagged as `large and extended'.
Note that the AT20G is known to be biased against extended sources due to a reduced sensitivity (50\% reduction in amplitude for source size $>45$~arcsec) and all these sources were targeted specifically by \cite{burke09}, see in particular their Table~2. Their total integrated flux measurements are reported in the catalogue we are using in this work, so that when we compare with the compact sources, we are not underestimating the flux densities of the former. In practical terms, we are using the flux coming from the innermost 1--2~kpc of the source (e.g. for Cen~A the scale is 0.037 kpc/arcsec, which means that the 20~GHz flux densities refer to the innermost 1.57~kpc, i.e. the core of the active galaxy and the initial part of the jets that extend for $\gsim$~1~Mpc). Also \cite{murphy10} comment further on this, in the context of associating the radio emission with the core of NGC~1068 (and other close galaxies), and point out that the reported flux appears to be associated with the active nucleus.

\section{Results}   
We explored the possible correlation between the hard X--ray flux (15--55 keV) and the radio flux density at 20GHz. Fig.~\ref{20ghz_flux} shows the scatter plot for the associations. There is no correlation between the two quantities for the Seyfert galaxies as also confirmed by the Spearman's rho correlation coefficient $r_s = 0.38$ (corresponding to a chance correlation probability $P= 0.11$). This correlation is even weaker if we include the blazars. Therefore we find no evidence of a relation between the fluxes. 

Note that four associations are Compton--thick sources: NGC~612 and NGC~1068 (both present in \citealp{burlon11}), NGC~4945, and Circinus. We show their absorbed and unabsorbed\footnote{For NGC~4945 and Circinus we estimated the suppression factor using the prescriptions of \cite{burlon11}, which are consistent with the findings of \cite{brightman11}. To this aim we used a fiducial value of $N_H$  taken from the most recent literature (\citealp{yaqoob12} and \citealp{dellaceca08}, respectively)} 
X--ray fluxes (connected by a vertical dashed line) in Fig.\ref{20ghz_flux}.}

We over--plotted in Fig.~\ref{20ghz_flux} the two limiting fluxes of both surveys as a grey shaded areas. The absence of points in the upper part of the plane does not seem affected by a simple selection effect, while the lower right one is likely affected by the BAT sensitivity. 

\begin{figure*}
\includegraphics[scale=0.75]{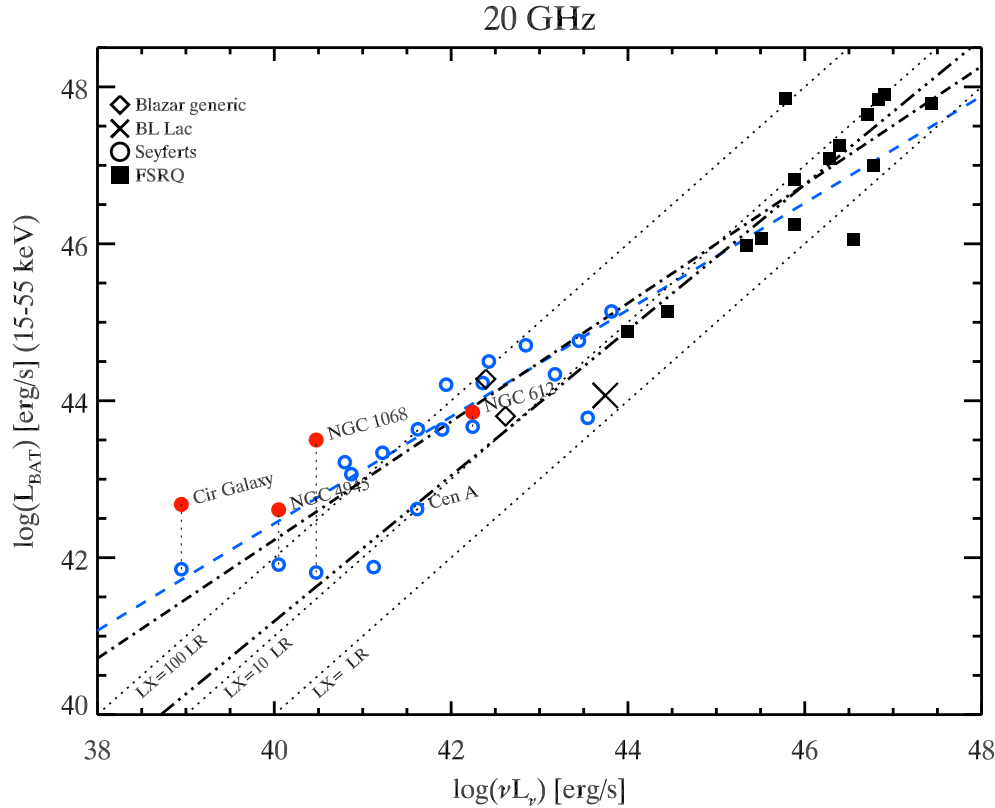}
{\caption{\sw/BAT luminosity vs. AT20G radio luminosity at 20~GHz. Different sources as per the legend and Fig.~\ref{20ghz_flux}. The best (bisector) fits are shown with a dashed blue, dash-dot, dash-triple-dot line for the Seyferts, all sample, and blazar respectively. The dotted lines represent three different $R_X$ values of 0.01, 0.1, 1 and are conversely expressed in the bottom left of the corner as relative radio--X luminosities. }\label{20ghz_lum} \vfil} 
\end{figure*}

However, the study of the correlation between the power output at different frequencies is often done in the luminosity space rather than in the flux space. 
We computed the $k$--corrected BAT luminosities via:
$$L_{\mathrm{BAT}} = 4\pi d_{\rm L}(z)^2 \frac{F_{15-55}}{(1+z)^{2-\Gamma}} $$
where $\Gamma$ is the photon index of the individual source fitted in the whole BAT energy band. Note that the usual notation of the X--ray photon indices is such that $F(\nu) \propto \nu^{1-\Gamma}$.
The $k$--corrected radio luminosity at the $i$--th central frequency (19.904~GHz, 8.64~GHz and 4.8~GHz) was obtained from the flux density $S(\nu)$ via:
$$L_{\mathrm{AT20}} = 4\pi d_{\rm L}(z)^2 \frac{\nu_i \times S(\nu_i)}{(1+z)^{1+\alpha}} $$
where $\alpha$ is the energy spectral index 
$(S(\nu) \propto \nu^{\alpha})$, 
so that we are calculating both luminosities in units of erg/s. We chose to use the spectral index between 5~GHz and 20~GHz as a proxy of the radio spectrum in the available ATCA band. We used the median spectral index for the sources which lacked the flux density measurement at 5~GHz.
A caveat for the more distant sources is that \cite{chhetri12} showed that the AT20G sources tend to show a turn over at the rest--frame frequency of about 30~GHz. This curvature effect does not significantly impact the $k$--corrected luminosities for all but the sources\footnote{For SWIFT~J1656.3--3302, SWIFT~J0539.9--2839, PKS~2126--158, QSO~B2052--47, PKS~0524--460 the radio luminosity is strictly speaking a lower limit.}
which are at high redshift ($z \gsim 1$) and harder than $\alpha=-1$.
 
The result is shown in Fig.~\ref{20ghz_lum}. The \lx--\lr\ correlation appears  stronger with respect to the flux plane, as can be seen in the corresponding Spearman's rho correlation coefficients and chance correlation probabilities (see the top panel of Table~\ref{tab:statistics}). 
A fit with the bisector method at 20~GHz yields 
$$logL_{X}= 0.68(\pm 0.11)\,logL_R + 15(\pm 5)$$
for the Seyfert sample, which is consistent with the slightly flatter result we obtain including also the blazars 
$logL_{X}=0.75(\pm 0.05)\,logL_R + 12(\pm 2)$.

\begin{table}
 \centering
  \caption{For each frequency block the first and second line are the Spearman's correlation coefficient and corresponding null hypothesis probability. The third and fourth lines show the corresponding partial correlation analysis results (labelled with a $(z)$), as described in the appendix.}
  \begin{tabular}{@{}l|c|c|c@{}}
  \hline
  \hline
    		&	All			&	Seyferts		&		Blazars\\
   \hline
20~GHz$^a$	&	(37)		&	(19)			&		(18)\\
	\hline
\lx--\lr\ 	&	0.94					&	0.86 				&	0.84\\	
			&	7.99$\times10^{-19}$	&	1.76$\times10^{-6}$	&	1.11$\times10^{-5}$\\
\lx--\lr\ (z) &	0.38					&	0.07				&	0.13\\	
			&	1.96	$\times10^{-2}$	&	0.78				&	0.59\\
	\hline		
8~GHz$^a$	&	(22)		&	(11)			&		(11)\\
	\hline      
\lx--\lr\ 	&	0.95					&	0.94					&	0.84	\\	
			&	6.74$\times10^{-12}$	&	1.12$\times10^{-5}$	&	1.33$\times10^{-3}$\\		
\lx--\lr\ (z) &	0.23			&	0.37			&	0.07\\	
			&	0.29			&	0.25			&	0.83\\
	\hline

5~GHz$^a$	&	(22)		&	(11)			&		(11)\\
	\hline      
\lx--\lr\ 	&	0.95					&	0.94				&	0.84	\\	
			&	8.55$\times10^{-12}$	&	1.12$\times10^{-5}$	&	1.33$\times10^{-3}$\\
\lx--\lr\ (z) &	0.26			&	0.42			&	0.05	 \\	
			&	0.23			&	0.19			&	0.87\\
	\hline		
	\hline
\end{tabular}
\begin{list}{}{}
\item[$^{\mathrm{a}}$] The numbers in parenthesis represent the number of AGN used to compute the statistics.
\end{list}
\label{tab:statistics}
\end{table}

However, the common dependence of the radio and X--ray luminosities with redshift should caution about the significance of the correlations. In order to remove the effect of $z$ we performed a partial correlation analysis, as detailed in the appendix.
 We find that the correlation is not significant for the sample of Seyfert--like AGN (chance correlation probability is 78\%). The apparent relation is therefore completely driven by their distance modulus. A caveat when considering the whole sample: from a purely statistical point of view the null hypothesis probability is $\sim 2 \times 10^{-2}$, corresponding to a confidence of 2.4\,$\sigma$. Nonetheless, since the redshift distribution is almost bimodal, we are effectively computing the partial correlation among two different populations. The Seyfert one with a lower median flux and distance versus the blazar one, with a much higher median flux and redshift. The results of the analysis in all the three available radio bands is summarised in Table~\ref{tab:statistics}.

\section{Discussion}

There is a natural scenario in which all X--ray emitting AGN are also radio emitters at some level, i.e. a scenario in which all AGN are able to accelerate a jet, even though only a small fraction ($\sim 10\%$, the genuine radio--loud) are able to successfully launch such a jet. In this ``aborted--jet'' scenario \citep{ghisellini04}, most AGN are indeed unable to launch the jet due to its subrelativisic speed, which make the jet collapse back and account for both radio and high--energy emission. The difference with the more well--known accretion scenario is due to the heating mechanism for the electrons responsible for the X--rays, which in the case of an aborted jet, is predicted to be localised along the (failed) jet axis. It is worth nothing that both scenarios are not necessarily in competition with each other, and may well be both contributing at different levels in the radio and high--energy bands. 

It is not surprising that lot of effort has been spent investigating the reciprocal contribution of radio emission and X--rays. In recent years numerous groups have tackled this issue from different perspectives. 
Different authors used soft X--ray \citep{canosa99, brinkmann00, salvato04}, and found a slope of the correlation which is close to unity\footnote{The slope is consistent with unity for \cite{brinkmann00} if the radio--quiet sample is considered.}. Some differences among the different works emerge as a consequence of the fitting method, and/or of the inclusion of some AGN which might be contaminated by a starburst component, and/or the definition of radio--loudness (as discussed before). \cite{canosa99} adopted monochromatic X--ray luminosities at 1~keV for a sample of $\sim 40$ ($z<0.1$) AGN. \cite{brinkmann00} chose 2~keV luminosities for $\sim 60$ ($z<1$) AGN , respectively. These were compared to the 5~GHz luminosity and a significant correlation was found (even if the effect of distance might have been underestimated). In the work of \cite{salvato04}, 53 local ($z < 0.1$) AGN were cross--correlated in the X--ray energy band of 0.5--2~keV versus 1.5~GHz monochromatic luminosities. 

Another difference emerges when we consider that the lower the radio frequency we adopt, the more we are contaminated by the lobes which reside further out. To the aim of checking the existence and validity of a relation between the jet output at radio and X--ray frequencies, shorter wavelengths are to be preferred. 
\cite{panessa07} studied an optically selected sample of $\sim$50 Seyfert galaxies, and compared the 2--10 keV luminosities to the 1.4, 5, and 15~GHz ones. In particular they found that the best fitting relation, taking into account also censored data, is represented by the same relation that holds for a sample of LLAGN, with a slope of between 0.71 and 0.97 at 5~GHz, and 0.8 and 1.02 at 15~GHz, with (without) the inclusion of a few radio--loud sources (according to the definition of \citealp{terashima03}) respectively. The difference between their Seyfert sample and the LLAGN is the mass range, which \cite{merloni03} showed to be the third parameter in the so--called ``fundamental plane of black holes'', which is anyhow only a scale parameter. It is worth noting that \cite{degasperin11} found only a mild relation in faint LLAGN between 2--10 keV X--rays and 4.8~GHz and 8.4~GHz.  
\begin{figure}
\includegraphics[width=\columnwidth]{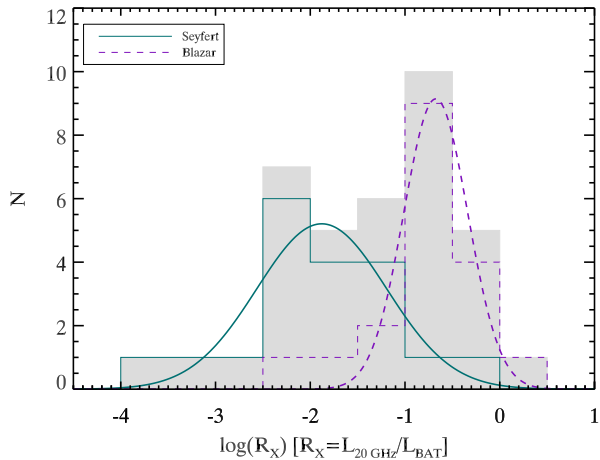}
\caption{Distributions of the $R_X$ parameter.} \label{rx}
\end{figure}
 
The framework of the ``fundamental plane of black holes'' is to establish both theoretically and observationally the coupling between the disk (i.e. the accretion power) and the jet. To that aim the accretion power linked to the X--ray emission from the corona and the radio power is associated to the jet. One of the result of their work is to successfully interpret their findings in the framework of the fundamental connection between a radiatively \emph{inefficient} accretion flow and the jet, under the assumption of scale invariance of the jet \citep{heinz03}. From an observational point of view they proved that this coupling holds at all scales (i.e. both for AGN and Galactic black holes). 

The absence of the radio--X correlation in our hard X--ray selected sample is at odds with the equivalent works at lower radio frequencies and lower X--ray energies. In principle this could be due to either an incompleteness of our sample (our result arises when cross--correlating $\sim 20$ AGN), or be physical in nature. 
We checked if the relation was weaker or stronger at different radio frequencies. Not all sources have counterparts in the AT20G at lower frequencies, but as we show in Table~\ref{tab:statistics} the result seems to hold at 8~GHz and 5~GHz. We looked at even lower frequencies,  using the fluxes from the NRAO VLA Sky Survey \citep[NVSS;][]{condon98},  Sydney University Molonglo Sky Survey \citep[SUMSS;][]{mauch03}, and MGPS-2 \citep{murphy07}, but unfortunately just 8 Seyfert AGN can be securely associated to our selection. These numbers are too low to run any meaningful statistical analysis. The problem of having too few counterparts to our cross--matched sample at lower frequencies, is mirrored at higher frequencies. We also checked the 30~GHz and 40~GHz flux densities from the Planck satellite Early Release Compact Source Catalog \citep{planck11} v--1.3\footnote{http://irsa.ipac.caltech.edu/Missions/planck.html}. Less than 5 Seyferts among our sources have high frequency measurements, making the higher frequency analysis even less significant.

Motivated by the similar result of absence of correlation when using the VLBI radio luminosities (Panessa \& Giroletti, 2013, MNRAS submitted), we run the following check on our sample. By using the definition of compactness adopted by \cite{chhetri12}, we excluded from our analysis the sources whose visibilities of the longest--over--shortest baselines were smaller than 0.86. This effectively translates in a factor of $\sim 10$ better angular resolution, down to 0.15 arcsec resolution\footnote{This translates into structures more compact than $\sim 9$ pc at the median Seyfert redshift, and $\sim 1.3$ kpc at the median blazar redshift}. 
Out of the whole sample, 21 AGN survive this selection and again the correlation is absent at the same statistical level to the original analysis.
Therefore we conclude that, if the absence of the correlation at high frequencies and high angular resolution will be proven statistically solid by future works, it could mean that the known radio--X correlation is relating a genuine ``core'' property (the X--rays) with a more distant one arising from the lobes of the jet. This in turn would imply that the X--ray output of  the AGN, averaged over the duration of the used survey, correlate with the energy reservoir represented by the distant radio structures.  

\emph{Beamed sources:} Note that \cite{merloni03} excluded the radio--loud (blazar--like) objects, and so do most authors. This is because in the blazar case both luminosities are associated to the relativistic jet and therefore the electromagnetic output is boosted by a factor $\propto \Gamma^2$. This is not normally the case for Seyfert galaxies, in which the jet --if present-- is assumed to be sub--relativistic. Testing the universality of the disk--jet coupling at yet other scales is beyond the scope of this paper. 
At the same time, they were motivated by the fact that the radio--X connection described by low mass systems and high mass systems defined parallel tracks in the luminosity plane. Indeed more massive objects like active galaxies tend to be more radio--loud with respect to the X--ray output when compared to Galactic black holes. The mass segregation and the accretion rate are in fact two important parameters of their treatment. It is nonetheless interesting to note a similar effect in our sample, if we compare the occupation of BAT selected Seyfert--like AGN and BAT selected FSRQ. In Fig.~\ref{20ghz_lum} we draw with dotted lines different radio--to--X ratios. Seyfert--like AGN tend to populate the area where the radio power is between 0.01 and 0.1 the X--ray power, while blazars tend to populate the more ``radio loud'' area where the ratio is between 0.1 and 1.

Another way to visualise this effect is to take the definition of radio loudness adopted by \cite{terashima03} adopting a similar ``high frequency'' radio loudness $R_X \equiv log(L_R/L_X)$. Fig.~\ref{rx} shows that the distribution of the $R_X$ parameter is different for the two classes of sources, which is a reflection of the two samples being displaced along different tracks. The median (and dispersion) of $log(R_X)$ for local Seyferts and blazars are  -1.88 (0.69), and -0.68 (0.34), respectively. The null hypothesis of the two populations of being drawn by the same parent sample has a probability $P=1.5 \times 10^{-4}$.

Our decision to include also the blazars in this work was to discuss the following: if the distribution of the Lorentz factors of the jets is peaked (i.e. if there is a ``typical'' $\Gamma$ for BAT FSRQ), then the beaming should impact the normalisation of the correlation drawn by their luminosities rather then the slope. Conversely, one has to consider that probably the dissipation of the jet power takes place in different areas along the jet, therefore the beaming will be different (higher) for the X--ray luminosities than for the radio luminosities. This effect likely affects also the slope of the correlation, but cannot be taken into account unless one independently knows the beaming factors for radio and X--rays. 
Even if beyond the goal of this paper, we would like to mention that if such a control on the Lorentz factors could be obtained (e.g. from variability in X--rays with the NuSTAR satellite \citep{harrison10}, and from VLBI in radio), then we would be able to de--boost the luminosities of blazars, and compare them properly with the Seyfert ones.

\section{Summary and conclusion}	
Starting from the six--year \sw/BAT survey of AGN, we looked for counterparts at high frequency radio in the AT20G survey \citep{murphy10} of the southern sky. By adopting a Bayesan cross--matching algorithm we found 37 AGN active in both regimes. Both surveys are biased towards bright objects, given that the sensitivity of the BAT is of the order of $6\times10^{-12}$ erg/cm$^2$/s and the AT20G has a cut at 40~mJy.
At the same time, the use of high energy X--rays is basically unbiased with respect to intervening gas, while the use of the highest radio frequency available helps in the contamination of the radio fluxes by extended structures. This is the best available set of observations (excluding VLBI) at these frequencies, in order to confront the ``core'' properties of AGN. 
We summarise our findings in the following:
\begin{itemize}
\item 	We investigated the radio--X connection at --so far-- unexplored X--ray frequencies \citep[but see][]{panessa11};
\item 	We found that only $\sim$20\% (considering that the AT20G covers the southern sky) of the hard X--ray AGN have a counterpart at 20~GHz, and these are equally distributed between local Seyfert galaxies and distant blazars;
\item 	The existence of a relation between the X--ray and radio output of Seyfert--like AGN in the form $L_X \propto L_R^{0.68}$ is essentially determined by the redshift;
\item	The revised radio loudness parameter shows a bimodality between the two classes of sources. A transition between the two could be ascribed to the fact that we are using a flux--limited sample, beamed luminosities for one class, and to a different mass distribution of these sources.
\end{itemize}

With the advent of the first focussed hard X--ray mission \cite[NuSTAR;][]{harrison10} there could be the interesting possibility to target the pre--selected sample of \sw/BAT AGN and get a lower limit on the Lorentz factor of the high--energy emission. From extensive VLBI measurements we would be able to resolve the very base also of Seyfert radio jets \cite[see e.g.][]{giroletti09} and even constrain the beaming factor of the radio luminosity. 
This will in principle allow the testing of the radio--X connection for both Seyfert and blazars at the same time, down to an hard X--ray sensitivity 100 times deeper than the current limit.



\section*{Acknowledgments}
DB is funded through ARC grant DP110102034. 
DB thanks R. Ekers and F. Panessa for fruitful discussions and useful feedback.
This research has made use of the NASA/IPAC Extragalactic Database (NED) which is operated by the Jet Propulsion Laboratory, California Institute of Technology, under contract with the National Aeronautics and Space Administration, and the SIMBAD database, operated at CDS, Strasbourg, France. Finally it has benefitted from extensive use of TOPCAT\footnote{http://www.star.bristol.ac.uk/$\sim$mbt/topcat/}.

\appendix
\section{Partial Correlation Analysis}
Given that both \lx\ and \lr\ have the same scaling with the distance modulus, we tested whether the redshift $z$ is driving the correlation \citep[see e.g.][]{padovani92}, as this is a well recognised problem when comparing quantities that depend on a third variable in the same fashion. To this aim we performed a  partial correlation analysis. Renaming the Pearson correlation coefficients 
$ r_{[L_{\mathrm{BAT}} L_{\mathrm{AT20}}]} = r_{12}$,
$ r_{[L_{\mathrm{BAT}} z]} = r_{13}$,
$ r_{[L_{\mathrm{AT20}} z]} = r_{23}$, and 
$ r_{[L_{\mathrm{BAT}} L_{\mathrm{AT20}} , z]} = r_{12,3}$
the first order correlation coefficient has the form \citep{kendall79}:
$$ r_{12,3} = \frac{r_{12} - r_{13}r_{23}} {\sqrt{(1-r_{13}^2)(1-r_{23}^2)}}$$

To test the significance of a non--zero value for $r$ we compute 
$$ t = \frac{r\sqrt{N-2}}{\sqrt{1-r^2}} $$
which obeys the probability distribution of the Student's $t$ statistic, with N-2 degrees of freedom, N being the number of elements of the samples we are testing. The corresponding probability of a chance correlation is computed numerically by integrating the area in the two tails of a Student's t--distribution function. The results are summarised in Table~\ref{tab:statistics}.
For each of the three frequencies we report in the first and second line the Spearman correlation coefficient and the corresponding chance correlation probability. In all frequencies and subsamples, this statistics alone seems to point to a very strong correlation between the observables (as often reported in the literature). In the third (labelled with a $z$ in parenthesis) and fourth line we report the first order Pearson partial correlation coefficient and the corresponding probability computed with a Student's t--distribution function (which can be converted into ``sigmas'' of confidence, should the reader prefer to).

\bibliographystyle{mn2e}


\label{lastpage}

\end{document}